# ARABIC TEXT SUMMARIZATION BASED ON LATENT SEMANTIC ANALYSIS TO ENHANCE ARABIC DOCUMENTS CLUSTERING

Hanane Froud[1], Abdelmonaime Lachkar[1] and Said Alaoui Ouatik[2]

[1] L.S.I.S, E.N.S.A,University Sidi Mohamed Ben Abdellah (USMBA),Fez, Morocco
hanane_froud@yahoo.fr, abdelmonaime_lachkar@yahoo.fr
[2] L.I.M, Faculty of Science Dhar EL Mahraz (FSDM),Fez, Morocco
s_ouatik@yahoo.com

## ABSTRACT

*Arabic Documents Clustering is an important task for obtaining good results with the traditional Information Retrieval (IR) systems especially with the rapid growth of the number of online documents present in Arabic language. Documents clustering aim to automatically group similar documents in one cluster using different similarity/distance measures. This task is often affected by the documents length, useful information on the documents is often accompanied by a large amount of noise, and therefore it is necessary to eliminate this noise while keeping useful information to boost the performance of Documents clustering. In this paper, we propose to evaluate the impact of text summarization using the Latent Semantic Analysis Model on Arabic Documents Clustering in order to solve problems cited above, using five similarity/distance measures: Euclidean Distance, Cosine Similarity, Jaccard Coefficient, Pearson Correlation Coefficient and Averaged Kullback-Leibler Divergence, for two times: without and with stemming. Our experimental results indicate that our proposed approach effectively solves the problems of noisy information and documents length, and thus significantly improve the clustering performance.*

## KEYWORDS

*Information Retrieval Systems, Arabic Language, Arabic Text Clustering, Arabic Text Summarization, Similarity Measures, Latent Semantic Analysis, Root and Light Stemmers.*

## 1. INTRODUCTION

There are several research projects investigating and exploring the techniques in traditional Information Retrieval (IR) systems for the English and European languages such as French, German, and Spanish and in Asian languages such as Chinese and Japanese. However, in Arabic language, there is little ongoing research in Arabic traditional Information Retrieval (IR) systems.

Moreover, the traditional Information Retrieval (IR) systems (without documents clustering) are becoming more and more insufficient for handling huge volumes of relevant texts documents, because to retrieve the documents of interest, the user must formulate the query using the keywords that appear in the documents. This is a difficult task for ordinary people who are not familiar with the vocabulary of the data corpus. Documents clustering may be useful as a complement to these traditional Information Retrieval (IR) systems, by organizing these documents by topics (clusters) in the documents feature space. It has been proved by Bellot & El-Bèze in [1] that document clustering increase the precision in Information Retrieval (IR) systems for French language.

On the other hand, for the Arabic Language Sameh H. Ghwanmeh in [2] presented a comparison study between the traditional Information Retrieval system and the clustered one. The concept of clustering documents has shown significant results on precision compared with traditional Information Retrieval systems without clustering. These results assure the results obtained by Bellot & El-Bèze [1] during their test on Amaryllis'99 corpora for French language.

Traditional documents clustering algorithms use the full-text in the documents to generate feature vectors. Such methods often produce unsatisfactory results because there is much noisy information in documents. The varying-length problem of the documents is also a significant negative factor affecting the performance. In this paper, we propose to investigate the use of summarization techniques to tackle these issues when clustering documents [13].

The goal of a summary is to produce a short representation of a long document. This problem can be solved by building an abstract representation of the whole document and then generating a shorter text or by selecting a few relevant sentences of the original text. With a large volume of text documents, presenting the user with a summary of each document greatly facilitates the task of finding the desired documents so:

- Text Summarization can be used to save time.
- Text Summarization can speed up other information retrieval and text mining processes.

In this paper, we propose to use the Latent Semantic Analysis to produce the Arabic summaries that we utilize to represent the documents in the Vector Space Model (VSM) and cluster them, in order to enhance the Arabic documents clustering [14].

Latent Semantics Analysis (LSA) has been successfully applied to information retrieval [13] [15][16][17] as well as many other related domains. It is based on Singular Value Decomposition (SVD), a mathematical matrix decomposition technique closely akin to factor analysis that is applicable to text corpora. Recently, LSA has been introduced into generic text summarization by [18].

This paper is organized as follows. The next section describes the Arabic summarization based Latent Semantic Analysis Model. Section 3 and 4 discuss respectively the Arabic text preprocessing, document representation used in the experiments, and the similarity measures. Section 5 explains experiment settings, dataset, evaluation approaches, results and analysis. Section6 concludes and discusses future work.

## 2. ARABIC TEXT SUMMARIZATION BASED ON LATENT SEMANTIC ANALYSIS MODEL

### 2.1. LSA Summarization

In this work, we propose to apply the Latent Semantic Analysis Model in order to generic Arabic Text Summarization [13] [17][18][19]. The process starts with the creation of terms by sentences matrix $A = [A_1 \; A_2 \; ... \; A_n]$ with each column vector $A_i$ representing the weighted term-frequency vector of sentence **i** in the document under consideration. The weighted term-frequency vector $Ai = [a_{1i} \; a_{2i} \; ... \; a_{ni}]^T$ of sentence i is defined as:

$$a_{ij} = L(t_{ij}).G(t_{ij})$$

where :
1. $L(t_{ji})$ is the local weighting for term j in sentence **i**: $L(t_{ji})=tf(t_{ji})$ where $tf(t_{ji})$ is the number of times term j occurs in the sentence.

2. G($t_{ji}$) is the global weighting for term j in the whole document: $G(t_{ij}) = \log(N / n(t_{ij}))$ where N is the total number of sentences in the document, and n($t_{ij}$) is the number of sentences that contain term j.

If there are a total of m terms and n sentences in the document, then we will have an m x n matrix A for the document.

Given an m x n matrix A (such as m≥n) the SVD of A is defined as [20]:

$$A = U \Sigma V^T$$

where U = [$u_{ij}$] is an m × n column-orthonormal matrix whose columns are called left singular vectors; Σ = diag($\sigma_1, \sigma_2, \ldots, \sigma_n$) is an n × n diagonal matrix, whose diagonal elements are non-negative singular values sorted in descending order, and V = [$v_{ij}$] is an n × n orthonormal matrix, whose columns are called right singular vectors. If rank(A) = r, then [21] Σ satisfies:

$$\sigma_1 \geq \sigma_2 \ldots \geq \sigma_r \succ \sigma_{r+1} = \ldots = \sigma_n = 0$$

The interpretation of applying the SVD to the terms by sentences matrix A can be made from two different viewpoints. From transformation point of view, the SVD derives a mapping between the m-dimensional space spawned by the weighted term-frequency vectors and the r-dimensional singular vector space. From semantic point of view, the SVD derives the latent semantic structure from the document represented by matrix A. This operation reflects a breakdown of the original document into r linearly-independent base vectors or concepts. Each term and sentence from the document is jointly indexed by these base vectors/concepts.

A unique SVD feature is that it is capable of capturing and modeling interrelationships among terms so that it can semantically cluster terms and sentences. Further-more, as demonstrated in [21], if a word combination pattern is salient and recurring in document, this pattern will be captured and represented by one of the singular vectors. The magnitude of the corresponding singular value indicates the importance degree of this pattern within the document. Any sentences containing this word combination pattern will be projected along this singular vector, and the sentence that best represents this pattern will have the largest index value with this vector. As each particular word combination pattern describes a certain topic/concept in the document, the facts described above naturally lead to the hypothesis that each singular vector represents a salient topic/concept of the document, and the magnitude of its corresponding singular value represents the degree of importance of the salient topic/concept.

Based on the above discussion, authors [18] proposed a summarization method which uses the matrix $V^T$. This matrix describes an importance degree of each topic in each sentence. The summarization process chooses the most informative sentence for each topic. It means that the k'th sentence we choose has the largest index value in k'th right singular vector in matrix $V^T$.

The proposed method in [18] is as follows:

1. Decompose the document D into individual sentences, and use these sentences to form the candidate sentence set S, and set k = 1.
2. Construct the terms by sentences matrix A for the document D.

3. Perform the SVD on A to obtain the singular value matrix $\Sigma$, and the right singular vector matrix $V^T$. In the singular vector space, each sentence **i** is represented by the column vector $\Psi_i = [v_{i1} v_{i2} ... v_{ir}]^T$ of $V^T$.

4. Select the k'th right singular vector from matrix $V^T$.

5. Select the sentence which has the largest index value with the k'th right singular vector, and include it in the summary.

6. If k reaches the predefined number, terminate the operation; otherwise, increment k by one, and go to Step 4.

In Step 5 of the above operation, finding the sentence that has the largest index value with the k'th right singular vector is equivalent to finding the column vector $\Psi_i$ whose k'th element $v_{ik}$ is the largest.

### 2.2. Arabic Summarization

In this paper we propose to use the above method to identify semantically important sentences for Arabic Summary creations (Figure 1) in order to enhance the Arabic Documents Clustering task.

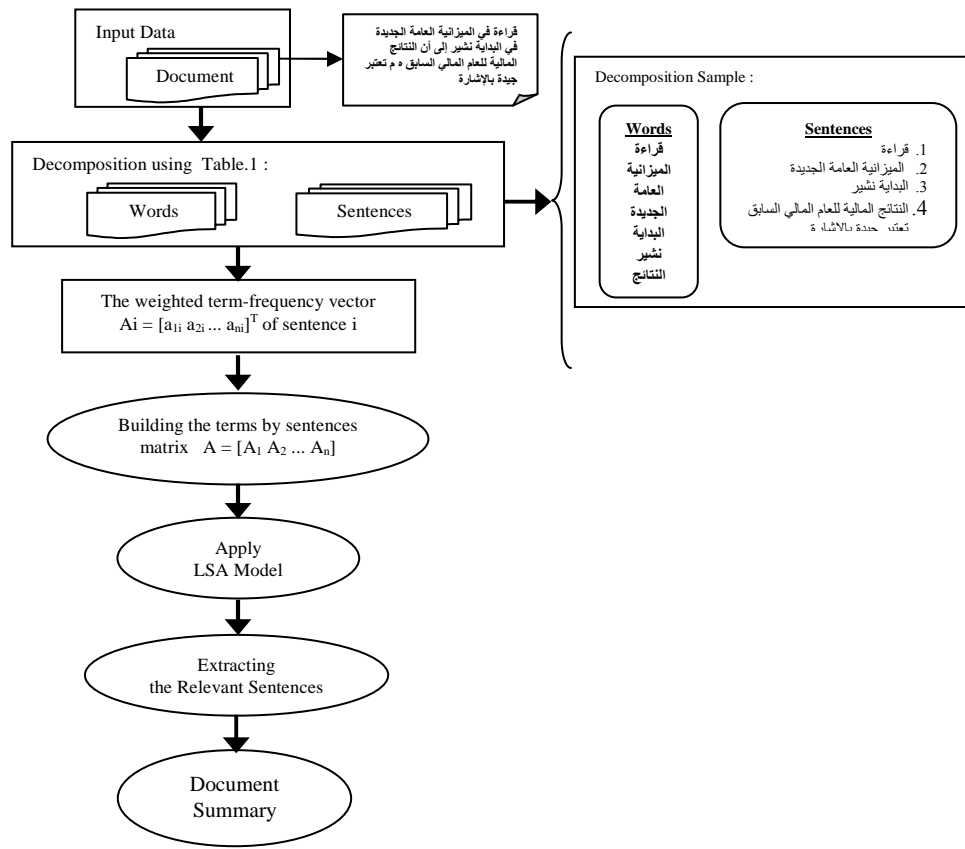

Figure 1. Arabic Text Summarization based on Latent Semantic Analysis Model

After building the test corpus, we decompose each document into individual sentences; this decomposition is a source of ambiguity, because on the one hand punctuation is rarely used in Arabic texts and other punctuation that, when it exists, is not always critical to guide the decomposition. In addition, some words can mark the beginning of a new sentence (or proposition).

For text decomposition [22] uses:

- ✓ A morphological decomposition based on punctuation,

- ✓ Decomposition based on the recognition of markers morphosyntactic or functional words such as: حتى, لكن, و, أو, or, and, but, when. However, these particles may play a role other than to separate phrases.

In our experiments, we use the morphosyntactic markers or functional words cited in [23] to decompose the document into individual sentences, in the following table we present some examples of these markers or functional words:

Table 1. Samples of Arabic Morphosyntactic Markers and Functional Words

| The Arabic Morphosyntactic Markers and Functional Words | أدوات الربط والوصل في اللغة العربية |
|---|---|
| in, and, then, or, but, when | في, و, ثم, أو, أم, بل, لكن, لكنْ, حتّى |
| also, after, although, as before, but this, not | أيضا, بعد, بالرغم, حيث, قبل, ولهذا, وليس |

## 3. ARABIC TEXT PREPROCESSING

### 3.1. Arabic Language Structure

The Arabic language is the language of the Holy Quran. It is one of the six official languages of the United Nations and the mother tongue of approximately 300 million people. It is a Semitic language with 28 alphabet letters. His writing orientation is from right-to-left. It can be classified into three types: Classical Arabic (العربية الفصحى), Modern Standard Arabic (العربية الحديثة) and Colloquial Arabic dialects (العربية العامية).

Classical Arabic is fully vowelized and it is the language of the holy Quran. Modern Standard Arabic is the official language throughout the Arab world. It is used in official documents, newspapers and magazines, in educational fields and for communication between Arabs of different nationalities. Colloquial Arabic dialects, on the other hand, are the languages spoken in the different Arab countries; the spoken forms of Arabic vary widely and each Arab country has its own dialect.

Modern Standard Arabic has a rich morphology, based on consonantal roots, which depends on vowel changes and in some cases consonantal insertions and deletions to create inflections and derivations which make morphological analysis a very complex task [24]. There is no capitalization in Arabic, which makes it hard to identify proper names, acronyms, and abbreviations.

### 3.2. Stemming

Arabic word Stemming is a technique that aim to find the lexical root or stem (Figure 2) for words in natural language, by removing affixes attached to its root, because an Arabic word can have a more complicated form with those affixes. An Arabic word can represent a phrase in English, for example the word ليحــدثونهم:"**to speak with them**" is decomposed as follows (Table 2):

Table 2. Arabic Word Decomposition

| Antefix | Prefix | Root | Suffix | Postfix |
|---|---|---|---|---|
| ل | ي | حدث | ون | هم |
| Preposition meaning "to" | A letter meaning the tense and the person of conjugation | speak | Termination of conjugation | A pronoun Meaning "them" |

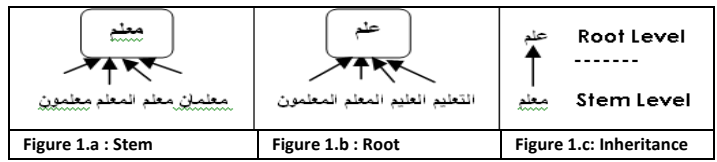

Figure 2. An Example of Root/Stem Preprocessing.

## 3.3. Root-based versus Stem-based approaches

Arabic stemming algorithms can be classified, according to the desired level of analysis, as *root-based approach* (Khoja [4]); and *stem-based approach* (Larkey [5]). In this section, a brief review on the two stemming approaches for stemming Arabic Text is presented.

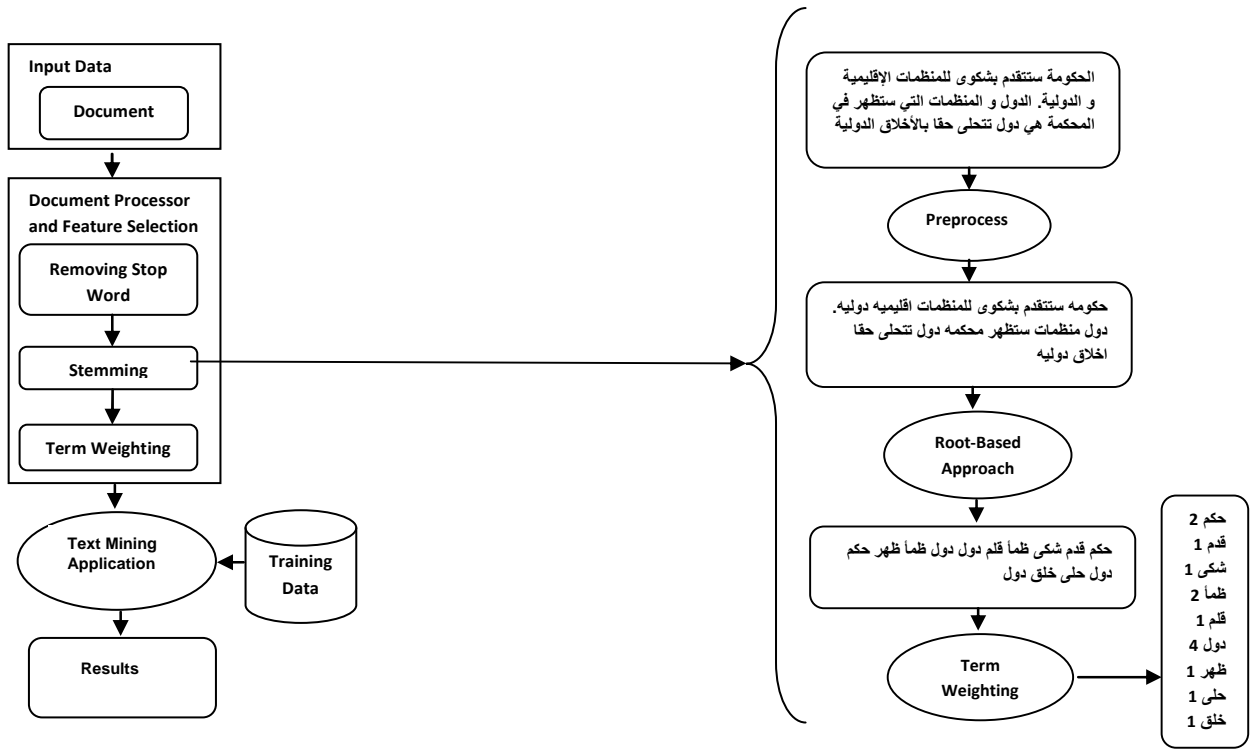

Figure 3. Example of Preprocessing with Khoja Stemmer algorithm

**Root-Based approach** uses morphological analysis to extract the root of a given Arabic word. Many algorithms have been developed for this approach. Al-Fedaghi and Al-Anzi algorithms try to find the root of the word by matching the word with all possible patterns with all possible affixes attached to it [25]. The algorithms do not remove any prefixes or suffixes. Al-Shalabi morphology system uses different algorithms to find the roots and patterns [26]. This algorithm removes the longest possible prefix, and then extracts the root by checking the first five letters of the word. This algorithm is based on an assumption that the root must appear in the first five letters of the word. Khoja has developed an algorithm that removes prefixes and suffixes, all the time checking that it's not removing part of the root and then matches the remaining word against the patterns of the same length to extract the root [4].

The aim of the **Stem-Based** approach or Light Stemmer approach is not to produce the root of a given Arabic word, rather is to remove the most frequent suffixes and prefixes. Light stemmer is mentioned by some authors [27,28,5,29], but till now there is almost no standard algorithm for Arabic light stemming, all trials in this field were a set of rules to strip off a small set of suffixes and prefixes, also there is no definite list of these strippable affixes.

In our work, we believe that the preprocessing of Arabic Documents is challenge and crucial stage. It may impact positively or negatively on the accuracy of any Text Mining tasks; therefore the choice of the preprocessing approaches will lead by necessity to the improvement of any Text Mining tasks very greatly.

To illustrate this, in Figure 2, we show an example using Khoja and Light stemmers. It produces different results: root and stem level related to the original word.

On the other hand Khoja stemmer can produce wrong results, for example, the word (منظمات) which means (organizations) is stemmed to (ظمأ) which means (he was thirsty) instead of the correct root (نظم).

Prior to applying document clustering techniques to an Arabic document, the latter is typically preprocessed: it is parsed, in order to remove stop words, and then words are stemmed using tow famous Stemming algorithms: **the Morphological Analyzer from Khoja and Garside [4], and the Light Stemmer developed by Larkey [5].** In addition, at this stage in this work, we computed the term-document using $tfidf$ weighting scheme.

### 3.4. Document Representation

There are several ways to model a text document. For example, it can be represented as a bag of words, where words are assumed to appear independently and the order is immaterial. This model is widely used in information retrieval and text mining [6].

Each word corresponds to a dimension in the resulting data space and each document then becomes a vector consisting of non-negative values on each dimension. Let $D = \{d_1,...,d_n\}$ be a set of documents and $T = \{t_1,...,t_m\}$ the set of distinct terms occurring in D. A document is then represented as an m-dimensional vector $\vec{t_d}$. Let $tf(d,t)$ denote the frequency of term $t \in T$ in document $t \in D$. Then the vector representation of a document d is:

$$\vec{t_d} = (tf(d,t_1),...,tf(d,t_m))$$

Although more frequent words are assumed to be more important, this is not usually the case in practice (in the Arabic language words like إلى that means to and فــي that means in). In fact, more complicated strategies such as the $tfidf$ weighting scheme as described below is normally used instead. So we choose in this work to produce the $tfidf$ weighting for each term for the document representation.

In the practice terms those appear frequently in a small number of documents but rarely in the other documents tend to be more relevant and specific for that particular group of documents, and therefore more useful for finding similar documents. In order to capture these terms and reflect their importance, we transform the basic term frequencies $tf(d,t)$ into the *tfidf* (term frequency and inversed document frequency) weighting scheme. *Tfidf* weights the frequency of a term t in a document d with a factor that discounts its importance with its appearances in the whole document collection, which is defined as:

$$tfidf(d,t) = tf(d,t) \times \log(\frac{|D|}{df(t)})$$

Here $df(t)$ is the number of documents in which term t appears, |D| is the numbers of documents in the dataset. We use $w_{t,d}$ to denote the weight of term t in document d in the following sections.

## 4. SIMILARITY MEASURES

In this section we discuss the five similarity measures that were tested in [3], and we include these five measures in our work to effect the Arabic text document clustering.

### 4.1. Metric

Not every distance measure is a metric. To qualify as a metric, a measure d must satisfy the following four conditions. Let x and y be any two objects in a set and $d(x,y)$ be the distance between x and y.

1. The distance between any two points must be non-negative, that is, $d(x,y) \geq 0$.

2. The distance between two objects must be zero if and only if the two objects are identical, that is, $d(x,y) = 0$ if and only if $x = y$.

3. Distance must be symmetric, that is, distance from x to y is the same as the distance from y to x, i.e. $d(x,y) = d(y,x)$.

4. The measure must satisfy the triangle inequality, which is $d(x,z) \leq d(x,y) + d(y,z)$.

### 4.2. Euclidean Distance

Euclidean distance is widely used in clustering problems, including clustering text. It satisfies all the above four conditions and therefore is a true metric. It is also the default distance measure used with the K-means algorithm.

Measuring distance between text documents, given two documents $d_a$ and $d_b$ represented by their term vectors $\vec{t_a}$ and $\vec{t_b}$ respectively, the Euclidean distance of the two documents is defined as

$$D_E(\vec{t_a},\vec{t_b}) = (\sum_{t=1}^{m} |w_{t,a} - w_{t,b}|^2)^{1/2},$$

where the term set is $T = \{t_1,...,t_m\}$. As mentioned previously, we use the *tfidf* value as term weights, that is $w_{t,a} = tfidf(d_a, t)$.

### 4.3. Cosine Similarity

Cosine similarity is one of the most popular similarity measure applied to text documents, such as in numerous information retrieval applications [6] and clustering too [7]. Given two documents $\vec{t_a}$ and $\vec{t_b}$, their cosine similarity is:

$$SIM_C(\vec{t_a}, \vec{t_b}) = \frac{\vec{t_a} \cdot \vec{t_b}}{|\vec{t_a}| \times |\vec{t_b}|},$$

where $\vec{t_a}$ and $\vec{t_b}$ are m-dimensional vectors over the term set $T = \{t_1,...,t_m\}$. Each dimension represents a term with its weight in the document, which is non-negative. As a result, the cosine similarity is non-negative and bounded between $[0,1]$. An important property of the cosine similarity is its independence of document length. For example, combining two identical copies of a document d to get a new pseudo document $d_0$, the cosine similarity between d and $d_0$ is 1, which means that these two documents are regarded to be identical.

### 4.4. Jaccard Coefficient

The Jaccard coefficient, which is sometimes referred to as the Tanimoto coefficient, measures similarity as the intersection divided by the union of the objects. For text document, the Jaccard coefficient compares the sum weight of shared terms to the sum weight of terms that are present in either of the two documents but are not the shared terms. The formal definition is:

$$SIM_J(\vec{t_a}, \vec{t_b}) = \frac{\vec{t_a} \cdot \vec{t_b}}{|\vec{t_a}|^2 + |\vec{t_b}|^2 - \vec{t_a} \cdot \vec{t_b}}$$

The Jaccard coefficient is a similarity measure and ranges between 0 and 1. It is 1 when the $\vec{t_a} = \vec{t_b}$ and 0 when $\vec{t_a}$ and $\vec{t_b}$ are disjoint. The corresponding distance measure is $D_J = 1 - SIM_J$ and we will use $D_J$ instead in subsequent experiments.

### 5.5. Pearson Correlation Coefficient

Pearson's correlation coefficient is another measure of the extent to which two vectors are related. There are different forms of the Pearson correlation coefficient formula. Given the term set $T = \{t_1,...,t_m\}$, a commonly used form is

$$SIM_P(\vec{t_a}, \vec{t_b}) = \frac{m\sum_{t=1}^{m} w_{t,a} \times w_{t,b} - TF_a \times TF_b}{\sqrt{\left[m\sum_{t=1}^{m} w_{t,a}^2 - TF_a^2\right]\left[m\sum_{t=1}^{m} w_{t,b}^2 - TF_b^2\right]}}$$

where $TF_a = \sum_{t=1}^{m} w_{t,a}$ and $TF_b = \sum_{t=1}^{m} w_{t,b}$

This is also a similarity measure. However, unlike the other measures, it ranges from -1 to +1 and it is 1 when $\vec{t_a} = \vec{t_b}$. In subsequent experiments we use the corresponding distance measure, which is $D_P = 1 - SIM_P$ when $SIM_P \geq 0$ and $D_P = |SIM_P|$ when $SIM_P \prec 0$.

### 4.6. Averaged Kullback-Leibler Divergence

In information theory based clustering, a document is considered as a probability distribution of terms. The similarity of two documents is measured as the distance between the two corresponding probability distributions. The Kullback-Leibler divergence (KL divergence), also called the relative entropy, is a widely applied measure for evaluating the differences between two probability distributions.

Given two distributions P and Q, the KL divergence from distribution P to distribution Q is defined as

$$D_{KL}(P \| Q) = P \log(\frac{P}{Q})$$

In the document scenario, the divergence between two distributions of words is:

$$D_{KL}(\vec{t_a} \| \vec{t_b}) = \sum_{t=1}^{m} w_{t,a} \times \log(\frac{w_{t,a}}{w_{t,b}}).$$

However, unlike the previous measures, the KL divergence is not symmetric, i.e. $D_{KL}(P \| Q) \neq D_{KL}(Q \| P)$. Therefore it is not a true metric. As a result, we use the averaged KL divergence instead, which is defined as:

$$D_{AvgKL}(P \| Q) = \pi_1 D_{KL}(P \| M) + \pi_2 D_{KL}(Q \| M),$$

where $\pi_1 = \frac{P}{P+Q}, \pi_2 = \frac{Q}{P+Q}$ and $M = \pi_1 P + \pi_2 Q$. For documents, the averaged KL divergence can be computed with the following formula:

$$D_{AvgKL}(\vec{t_a} \| \vec{t_b}) = \sum_{t=1}^{m} (\pi_1 \times D(w_{t,a} \| w_t) + \pi_2 \times D(w_{t,b} \| w_t)),$$

where $\pi_1 = \frac{w_{t,a}}{w_{t,a} + w_{t,b}}, \pi_2 = \frac{w_{t,b}}{w_{t,a} + w_{t,b}},$ and $w_t = \pi_1 \times w_{t,a} + \pi_2 \times w_{t,b}$

The average weighting between two vectors ensures symmetry, that is, the divergence from document i to document j is the same as the divergence from document j to document i. The averaged KL divergence has recently been applied to clustering text documents, such as in the family of the Information Bottleneck clustering algorithms [8], to good effect.

### 5. EXPERIMENTS AND RESULTS

In our experiments (Figure 4), we used the K-means algorithm as document clustering method. It works with distance measures which basically aim to minimize the within-cluster distances. Therefore, similarity measures do not directly fit into the algorithm, because smaller values indicate dissimilarity. The Euclidean distance and the averaged KL divergence are distance measures, while the cosine similarity, Jaccard coefficient and Pearson coefficient are similarity measures. [3] applies a simple transformation to convert the similarity measure to distance values. Because both cosine similarity and Jaccard coefficient are bounded in $[0,1]$ and monotonic, we take $D = 1 - SIM$ as the corresponding distance value. For Pearson coefficient, which ranges from −1 to +1, we take $D = 1 - SIM$ when $SIM \geq 0$ and $D = |SIM|$ when $SIM \prec 0$. For the testing dataset, we experimented with different similarity measures for three times: without stemming, and with stemming using the Morphological Analyzer from Khoja and Garside [4], and the Light Stemmer [5], in two case: in the first one, we apply the proposed method above to summarize for the all documents in dataset and then cluster them. In the second case, we cluster the original documents without summarization. Moreover, each experiment was run 5 times and the results are the averaged value over 5 runs. Each run has different initial seed sets.

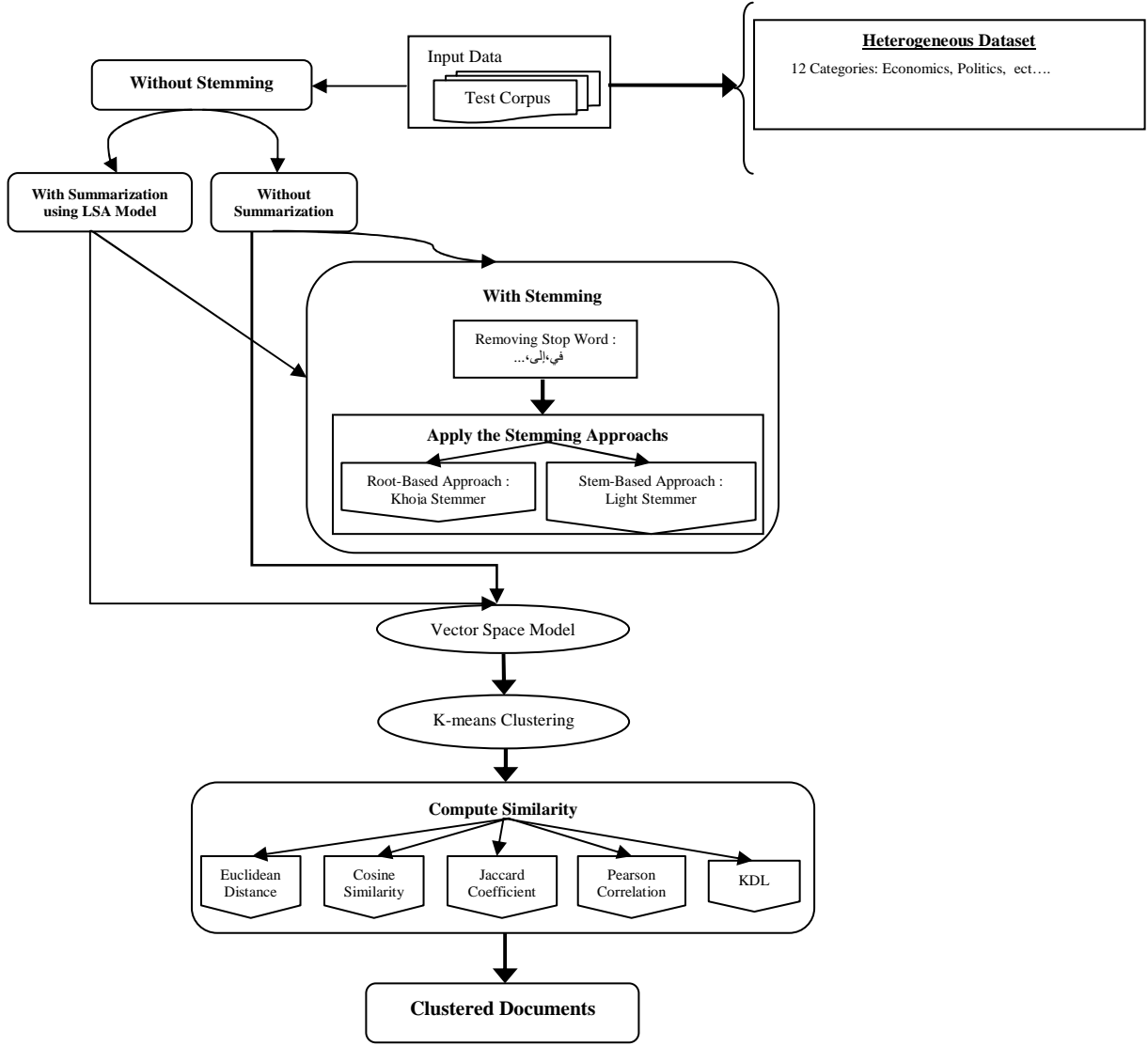

Figure 4. Description of Our Experiments

## 5.1. Dataset

The testing dataset [9] (Corpus of Contemporary Arabic (CCA)) is composed of 12 several categories, each latter contains documents from websites and from radio Qatar. A summary of the testing dataset is shown in Table 3.

As mentioned previously, the baseline method is the full-text representation, for each document, we removed stop words and stem the remaining words by using Khoja stemmer's and Larkey stemmer's. Then, to illustrate the benefits of our proposed approach, we use document summaries to cluster our dataset.

Table 3. Number of texts and number of Terms in each category of the testing dataset

| Text Categories | Number of Texts | Number of Terms |
|---|---|---|
| Economics | 29 | 67 478 |
| Education | 10 | 25 574 |
| Health and Medicine | 32 | 40 480 |
| Interviews | 24 | 58 408 |
| Politics | 9 | 46 291 |
| Recipes | 9 | 4 973 |
| Religion | 19 | 111 199 |
| Science | 45 | 104 795 |
| Sociology | 30 | 85 688 |
| Spoken | 7 | 5 605 |
| Sports | 3 | 8 290 |
| Tourist and Travel | 61 | 46 093 |

## 5.2. Results

The quality of the clustering result was evaluated using two evaluation measures: purity and entropy, which are widely used to evaluate the performance of unsupervised learning algorithms [10] [11].

The *purity* measure evaluates the coherence of a cluster, that is, the degree to which a cluster contains documents from a single category. Given a particular cluster $C_i$ of size $n_i$, the purity of $C_i$ is formally defined as:

$$P(C_i) = \frac{1}{n_i} \max_h (n_i^h)$$

where $\max_h(n_i^h)$ is the number of documents that are from the dominant category in cluster $C_i$ and $n_i^h$ represents the number of documents from cluster $C_i$ assigned to category h. In general, the higher the purity value, the better the quality of the cluster is.

The entropy measure evaluates the distribution of categories in a given cluster. The entropy of a cluster $C_i$ with size $n_i$ is defined to be

$$E(C_i) = -\frac{1}{\log c} \sum_{h=1}^{k} \frac{n_i^h}{n_i} \log(\frac{n_i^h}{n_i})$$

where c is the total number of categories in the data set and $n_i^h$ is the number of documents from the hth class that were assigned to cluster $C_i$.

The entropy measure is more comprehensive than purity because rather than just considering the number of objects in and not in the dominant category, it considers the overall distribution of all the categories in a given cluster. Contrary to the purity measure, for an ideal cluster with documents from only a single category, the entropy of the cluster will be 0. In general, the smaller the entropy value, the better the quality of the cluster is. Moreover, the averaged entropy of the overall solution is defined to be the weighted sum of the individual entropy value of each cluster, that is,

$$Entropy = \sum_{i=1}^{k} \frac{n_i}{n} E(C_i)$$

where n is the number of documents in our dataset.

In the following, The Table 4 and the Table 5 show the average purity and entropy results for each similarity/distance measure with the Morphological Analyzer from Khoja and Garside [4], the Larkey's Stemmer [5], and without stemming using the full- text representation.

On the other hand, the Table 6 and the Table 7 illustrate the results using document summaries with the same stemmers and similarity/distance measures.

### 5.2. 1. Results Using Full-Text Representation

*5.2. 1.a. Results with Stemming*

In Table 4, with Khoja's stemmer, the overall purity values for the Euclidean Distance, the Cosine Similarity and the averaged KL Divergence are quite similar and perform bad relatively to the other measures. Meanwhile, the Jaccard measure is the better in generating more coherent clusters with a considerable purity score.

In this context, using the Larkey's stemmer, the purity value of the averaged KL Divergence measure is the best one with only 1% difference relatively to the other four measures.

Table 4. Purity and Entropy Results with **Khoja's Stemmer**, and **Larkey's Stemmer** Using **Full-Text Representation**

|  |  | Euclidean | Cosine | Jaccard | Pearson | KLD |
|---|---|---|---|---|---|---|
| Khoja's stemmer | Entropy | 0.26 | **0.25** | **0.23** | 0.27 | 0.26 |
|  | Purity | 0.6 | 0.6 | **0.64** | 0.61 | 0.6 |
| Larkey's stemmer | Entropy | **0. 286** | 0. 286 | 0. 286 | 0. 286 | 0.30 |
|  | Purity | 0.52 | 0.52 | 0.52 | 0.52 | **0.53** |

*5.2. 2.b. Results without Stemming*

The Table 5, shows the higher purity scores (0.77) than those shown in the Table 4 for the Euclidean Distance, the Cosine Similarity and the Jaccard measures. In the other hand the Pearson Correlation and averaged KL Divergence are quite similar but still better than purity values for these measures in the Table 4.

The overall entropy value for each measure is shown in the two Tables. Again, the best results are there in the Table 5 that shows the better and similar entropy values for the Euclidean Distance, the Cosine Similarity and the Jaccard measures. However, the averaged KL Divergence performs worst than the other measures but better than the other one in the other Table (Table 4).

Table 5. Purity and Entropy Results **without Stemming** Using **Full-Text Representation**

|         | Euclidean | Cosine | Jaccard | Pearson | KLD  |
|---------|-----------|--------|---------|---------|------|
| Entropy | **0.16**  | **0.16** | **0.16** | 0.17   | 0.18 |
| Purity  | **0.77**  | **0.77** | **0.77** | 0.69   | 0.69 |

### 5.2. 2. Results Using Document Summaries

*5.2. 2.a. Results with Stemming*

Table 6 presents the average purity and entropy results for each similarity/distance measures using document summaries instead the full-text representation with Khoja's stemmer and Larkey's stemmer.

As shown in Table 6, for the two stemmers, Euclidean Distance, Cosine Similarity, and Jaccard measures are slightly better in generating more coherent clusters which means the clusters have higher purity and lower entropy scores. On the other hand, Pearson and KLD measures perform worst relatively to the other measures. Comparing these results with those obtained in Table 4, we can conclude that the obtained scores was improved specially the overall entropy values.

Table 6. Purity and Entropy Results with **Khoja's Stemmer**, and **Larkey's Stemmer** Using **Documents Summaries**

|                   |         | Euclidean  | Cosine     | Jaccard    | Pearson | KLD   |
|-------------------|---------|------------|------------|------------|---------|-------|
| Khoja's stemmer   | Entropy | **0.1272** | **0.1275** | **0.1275** | 0.150   | 0.151 |
|                   | Purity  | **0.385**  | **0.385**  | **0.385**  | 0.381   | 0.370 |
| Larkey's stemmer  | Entropy | **0.168**  | **0.169**  | **0.169**  | 0.173   | 0.178 |
|                   | Purity  | 0.336      | **0.342**  | 0.339      | 0.316   | **0.342** |

*5.2. 2.b. Results without Stemming*

A closer look at Tables 5 and 7 shows that, in this latter, the overall entropy values of Euclidean Distance, Cosine Similarity, Jaccard and Pearson measures are nearly similar and proves their ability to produce coherent clusters.

On the one side, in the Table 6 we can remark that the purity scores (**0.385** Khoja's stemmer, **0.339** Larkey's stemmer) **are** generally **higher** than those shown in the Table 7 for the all similarity/distance measures, on the other side, the overall entropy values in this table for the Euclidean Distance, the Cosine Similarity and the Jaccard measures with Khoja's stemmer performs bad than those in the Table 7. However, with Larkey's stemmer the overall entropy values for each measure performs contrary to their exiting in Table 7.

Table 7. Purity and Entropy Results **without Stemming** Using **Documents Summaries**

|         | Euclidean | Cosine    | Jaccard   | Pearson | KLD   |
|---------|-----------|-----------|-----------|---------|-------|
| Entropy | 0.154     | **0.153** | **0.152** | 0.154   | 0.181 |
| Purity  | **0.325** | 0.320     | 0.322     | 0.330   | 0.319 |

The above results lead as to conclude that:

First, the Tables 4 and 5 show that the use of stemming affects negatively the clustering, this is mainly due to **the ambiguity** created when we applied the stemming (for example, we can obtain two roots that made of the same letters but semantically different). Our observation broadly agrees with M.El kourdi, A.Bensaid, and T.Rachidi in [12], and with our works in [14][17].

Second, the obtained overall entropy values shown in Tables 6 and 7 proves that the summarizing documents can make their topics salient and improve the clustering performance [13] for two times: with and without stemming. However, the obtained purity values seem not promising to improve the clustering task; this is can be due to the bad choice of the number of sentences in summaries because this latter has great impact on the quality of summaries thus could lead to different clustering results. Too few sentences will result in mach sparse vector representation and are not enough to represent the document fully. Too many sentences may introduce noise and degrade the benefits of the summarization.

# 6. CONCLUSION

In this paper, we have proposed to illustrate the benefits of the summarization using the Latent Semantic Analysis Model, by comparing the clustering results based on summarization with the full-text baseline on the Arabic Documents Clustering for five similarity/distance measures for three times: without stemming, and with stemming using Khoja's stemmer, and the Larkey's stemmer.

We found that the Euclidean Distance, the Cosine Similarity and the Jaccard measures have comparable effectiveness for the partitional Arabic Documents Clustering task for finding more coherent clusters in case we didn't use the stemming for the full-text representation. On the other hand the Pearson Correlation and averaged KL Divergence are quite similar in theirs results but there are not better than the other measures in the same case.

Instead of using full-text as the representation for document clustering, we use LSA model as summarization techniques to eliminate the noise on the documents and select the most salient sentences to represent the original documents. Furthermore, summarization can help overcome the varying length problem of the diverse documents. In our experiments using document summaries, we remark that again the Euclidean Distance, the Cosine Similarity and the Jaccard measures have comparable effectiveness to produce more coherent clusters than the Pearson Correlation and averaged KL Divergence, in the two times: with and without stemming.

**Authors**

**Miss. Hanane Froud** Phd Student in Laboratory of Information Science and Systems, ECOLE NATIONALE DES SCIENCES APPLIQUÉES, University Sidi Mohamed Ben Abdellah (USMBA), Fez, Morocco. She has also presented different papers at different National and International conferences.

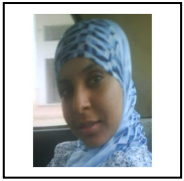

**Pr. Abdelmonaime LACHKAR** received his PhD degree from the USMBA, Morocco in 2004 in computer science; He is working as a Professor and Head of Computer Science and Engineering (E.N.S.A), in University Sidi Mohamed Ben Abdellah (USMBA), Fez, Morocco. His current research interests include Arabic Text Mining Applications: Arabic Web Document Clustering and Categorization. Arabic Information and Retrieval Systems, Arabic Text Summarization, etc …, Image Indexing and Retrieval, 3D Shape Indexing and Retrieval in large 3D Objects Databases, Color Image Segmentation, Unsupervised clustering, Cluster Validity Index, on-line and off-line Arabic and Latin handwritten recognition, and Medical Image Applications.

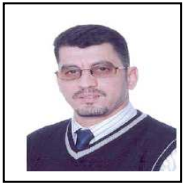

**Pr. Said Alaoui Ouatik** is working as a Professor in Department of Computer Science, Faculty of Science Dhar EL Mahraz (FSDM), Fez, Morocco. His research interests include high-dimensional indexing and content-based retrieval, Arabic Document Categorization. 2D/3D Shapes Indexing and Retrieval in large 3D Objects Database.

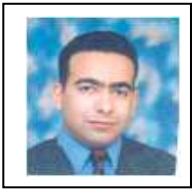